# Chiral Tetrahedrons as Unitary Quaternions: Molecules and Particles Under the Same Standard?

S. CAPOZZIELLO,[a,]* A. LATTANZI[b,]*

[a]Dipartimento di Scienze Fisiche and INFN (sez. di Napoli), Università di Napoli "Federico II", Complesso Universitario di Monte S. Angelo, Via Cinthia, I-80126, Napoli, Italy

[b]Dipartimento di Chimica, Universita' di Salerno, Via S. Allende, I- 84081, Baronissi, Salerno, Italy

capozziello@sa.infn.it; lattanzi@unisa.it

**ABSTRACT:** Chiral tetrahedral molecules can be dealt under the standard of quaternionic algebra. Specifically, non-commutativity of quaternions is a feature directly related to the chirality of molecules. It is shown that a quaternionic representation naturally comes out from empirical Fischer projections and the action of the related algebra on the quantum states of molecules is studied. Analogies with fundamental particle physics are discussed.

## 1. Introduction

The use of analogies often gives deep insight to understand phenomena which appear very different in length, time and energy scales. For example, Heisenberg [1] suggested that the analogies between quantum states of chiral molecules and those of elementary particles have to be seriously taken into account, in order to develop a coherent quantum chemistry capable of providing a comprehensive description of structures and dynamics of molecules. In this perspective, the representations adopted for particle quantum mechanics, in particular the group theory, have to work in the framework of quantum chemistry, giving account of most of the features of particles.

Recently, it has been shown that chiral tetrahedral molecules can be represented by an orthogonal $O(4)$ algebra, constructed starting from empirical Fischer projections [2,3]. This representation gives rise to a chirality index $\chi$ which allows the classification of molecules as achiral, diastereoisomers and enantiomers. As a consequence, the chiral features of tetrahedral chains can be predicted by means of a sort of molecular Aufbau [4]. Furthermore, a consistent Schrödinger equation has been worked out, whose solutions represent the bonds of tetrahedrons and then the $O(4)$ algebra can be considered as a "quantum chiral algebra" [5].

This approach can be furtherly developed taking into account the quaternionic representation which holds a prominent role in fundamental particle physics. In fact, it is well-known that Dirac equation proceeds from the Klein-Gordon equation, when written in a quaternionic form [6,7], and Pauli matrices satisfy quaternionic algebraic rules [8]. These are the fundamental features of spinor particle physics, so that one can wonder if they also hold for chiral molecules. Specifically, the non-commutativity of quaternionic algebra can be directly related to chirality and parity of particles and particle quantum states can be represented as quaternionic vectors.

In this paper, we want to deal with tetrahedral molecules as quaternions, showing that this is a natural representation by which chirality is intrinsically recovered. On the other hand, it follows that the $O(4)$ chiral algebra is nothing else but a quaternionic representation for non-commutative objects.

The layout of the paper is the following. In Section 2, the main features of quaternions, as generalized complex numbers and vectors, are described. In particular, we deal with their $SO(4)$ (and then $O(4)$) algebraic representation. Section 3 is devoted to the construction of quaternionic representation of chiral molecules starting from Fischer projection rules. There will be given account of eigenvalues, eigenvectors and commutation rules. The action of such a quantum algebra on the quantum states of molecules is discussed in Section 4, pointing out how the approach could be of effective interest for quantum chemistry. Discussion and conclusions, in particular analogies with elementary particles, are reported in Section 5.

## 2. Properties and representations of quaternions

Quaternions are generalized complex numbers given by the linear combinations

$$q = a + ib + jc + kd, \quad q \in H \tag{1}$$

where a, b, c, d are real coefficients, i, j, k are unitary immaginary numbers and H is the set of all quaternions. Let us introduce the bilinear multiplication rules in H:

$$ij = k = -ji, \quad jk = i = -kj, \quad ki = j = -ik, \tag{2}$$

$$i^2 = j^2 = k^2 = -1, \tag{3}$$

which can be expressed as anticommutators being

$$ij + ji = \{i, j\} = 2k, \quad jk + kj = \{j, k\} = 2i, \quad ki + ik = \{k, i\} = 2j \tag{4}$$

This is an associative, non-commutative algebra which can be represented in matrix form. The quaternion q can be given as a matrix $A(q) \in M(2, C)$ where

$$A(q) = \begin{pmatrix} a + ib & c + di \\ -c + id & a - ib \end{pmatrix}, \tag{5}$$

where i is the imaginary unit. Given the multiplication property

$$A(q_1 q_2) = A(q_1) A(q_2), \tag{6}$$

it means that the application $q \rightarrow A(q)$ is a homomorphism.

If $q = i, j, k,$ from Eq. (6), it follows

$$A(i) = \begin{pmatrix} i & 0 \\ 0 & -i \end{pmatrix} \quad A(j) = \begin{pmatrix} 0 & 1 \\ -1 & 0 \end{pmatrix} \quad A(k) = \begin{pmatrix} 0 & i \\ i & 0 \end{pmatrix}, \tag{7}$$

and then

$$A(i)A(j) = A(k) \tag{8}$$

and analogue equations. Matrices A can be represented as

$$\sigma_x = -iA(k), \quad \sigma_y = -iA(j), \quad \sigma_z = -iA(i), \tag{9}$$

which are nothing else but the "Pauli matrices" which satisfy the properties

$$\sigma_x^2 = \sigma_y^2 = \sigma_z^2 = 1, \quad \sigma_x \sigma_y = -\sigma_y \sigma_x = i\sigma_z, \tag{10}$$

and analogues. This is another form for the relations (2), (3), (4). As it is well-known, the Pauli matrices constitute a $SU(2)$ algebra which is fundamental in spinor particle physics.

Let us now introduce the conjugation operation in H, which is

$$\bar{q} = a - ib - jc - kd. \tag{11}$$

From the above definitions, we have

$$A(\bar{q}) = \overline{A}^T(q), \tag{12}$$

and then the matrices A and $\overline{A}$ are orthogonal. This means that

$$AA^T = \overline{AA}^T = (a^2 + b^2 + c^2 + d^2)\mathbf{I} \tag{13}$$

where $\mathbf{I}$ is the identity matrix and the norm of a quaternion is given by

$$|q|^2 = q\bar{q} = a^2 + b^2 + c^2 + d^2 = \det A(q). \tag{14}$$

Any quaternion q, different from zero with $|q|^2 \neq 0$, has an inverse $q^{-1}$ so that

$$qq^{-1} = 1, \tag{15}$$

and then

$$q^{-1} = \frac{\bar{q}}{|q|^2}. \tag{16}$$

A particularly useful subset of quaternions is that with unitary norm, usually denoted as $H_1$. In this case, the property

$$q^{-1} = \bar{q}, \quad \forall q \in H_1 \tag{17}$$

holds. In a 4-dimensional real space $R^4$, with coordinates (a, b, c, d), the set $H_1$ is a hypersurface whose equation is

$$a^2 + b^2 + c^2 + d^2 = 1 \tag{18}$$

that is a 3-dimensional unitary sphere in $R^4$.

Considering a unit quaternion $q = a + ib + jc + kd$, $|q|^2 = 1$ and $x = a + ib$, $y = c + id$, we have $|x|^2 + |y|^2 = 1$ and then

$$A(q) = \begin{pmatrix} a + ib & c + di \\ -c + id & a - ib \end{pmatrix} = \begin{pmatrix} x & y \\ -\overline{y} & \overline{x} \end{pmatrix} \tag{19}$$

which means that the group $H_1$ is isomorphic to $SU(2)$.

If $SO(3)$ is the group of rotations on the sphere, in standard 3-dimensional Euclidean space $R^3$, it is straightforward to recover the isomorphism [9]

$$SO(3) \approx SU(2)/\pm 1 \tag{20}$$

so that the unitary quaternions can represent rotations using the Pauli matrices.

Another useful representation is given by the isomorphism of the groups

$$SU(2) \times SU(2)/\pm 1 \approx SO(4). \tag{21}$$

If $p, q \in H_1 \approx SU(2)$, the application

$$\Lambda_{p,q} : X \to pXq^{-1}, \quad X \in H \approx R^4 \tag{22}$$

preserves the squared norm of the quaternion X and it is an orthogonal transformation in the 4-dimensional space H. Clearly, $\Lambda_{-p,-q} = \Lambda_{p,q}$ and this is nothing else but a rotation in $R^4$. Including inversions, we recover the $O(4)$ algebra starting from unitary quaternions.

Finally, quaternions can be represented as 4-vectors

$$q = \begin{pmatrix} a \\ b \\ c \\ d \end{pmatrix}, \quad p = \begin{pmatrix} a_1 \\ b_1 \\ c_1 \\ d_1 \end{pmatrix}. \tag{23}$$

Their multiplication can be given as matrix transformations, which are, using the rules (2), (3),

$$pq = \begin{pmatrix} a & -b & -c & -d \\ b & a & -d & c \\ c & d & a & -b \\ d & -c & b & a \end{pmatrix} \begin{pmatrix} a_1 \\ b_1 \\ c_1 \\ d_1 \end{pmatrix} \equiv Pq \quad (24)$$

and, being non-commutative,

$$qp = \begin{pmatrix} a & -b & -c & -d \\ b & a & d & -c \\ c & -d & a & b \\ d & -c & -b & a \end{pmatrix} \begin{pmatrix} a_1 \\ b_1 \\ c_1 \\ d_1 \end{pmatrix} \equiv \overline{P}q. \quad (25)$$

The matrices $P$ and $\overline{P}$ are orthogonal and, from Eq. (13), it is

$$PP^T = \overline{P}\overline{P}^T = (a^2 + b^2 + c^2 + d^2)\mathbf{I} \quad (26)$$

Orthogonal matrices in 4-dimensions constitute the $O(4)$ group which is given by the group of 4-rotations $SO(4)$ plus the inversions which do not constitute a group since the product of two inversions is again a rotation. As we shall see in the next section, this representation works for tetrahedral molecules with central molecular chirality.

## 3. Chiral tetrahedral molecules as quaternions

A chiral tetrahedral molecule can be described as a colum vector (a quaternion)

$$\Psi = \begin{pmatrix} \Psi_1 \\ \Psi_2 \\ \Psi_3 \\ \Psi_4 \end{pmatrix} \quad (27)$$

where the components $\Psi_j$ represent the bonds connected together by the central atom. Starting from a given configuration, all possible rotations and permutations of such bonds can be obtained by the Fischer projections [10], which assign the planar representation of the molecule. The rules which we have to follow for this representation can be sketched starting from Fig.1: the atoms pointing sideways must project forward in the model, while those pointing up and down in the projection must extend toward the rear. Let us take into account $(S)$-$(+)$-lactic acid.

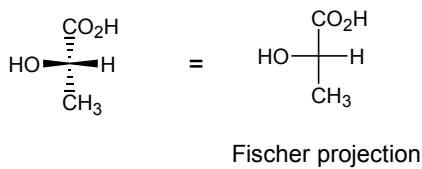

Fischer projection

**Figure 1.** Fischer projection of (*S*)-(+)-lactic acid.

To obtain proper results, projections may not be rotated of 90°, while a 180° rotation is allowed. The interchange of any two groups results in the conversion of an enantiomer into its mirror image (Fig.2).

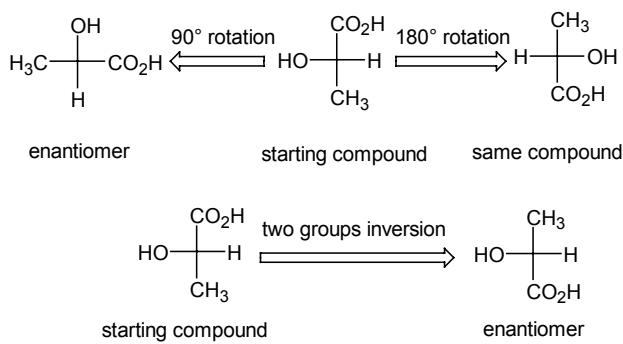

**Figure 2.** Fundamental rules to handle Fischer projections.

The chemical groups can be indicated by numbers running from 1 to 4. For the example which we are considering: OH=1, CO$_2$H=2, H=3, CH$_3$=4, without taking into account the effective priorities of the groups [10]. There are 24 (=4! the number of permutations of 4 ligands among 4 sites) projections.

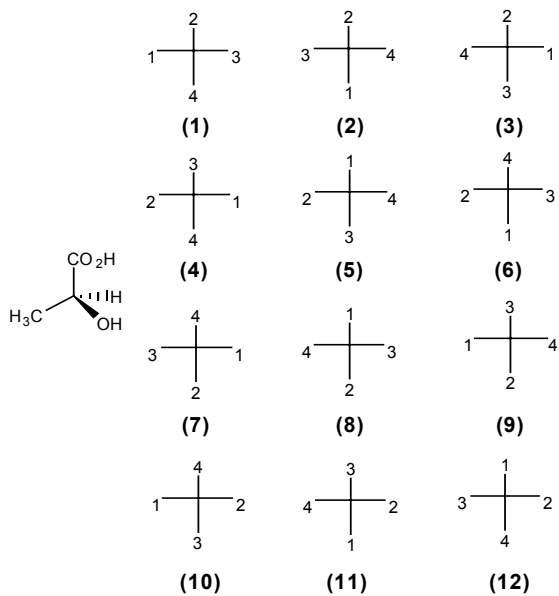

**Figure 3.** Twelve Fischer projections of (*S*)-(+)-lactic acid.

Twelve of these correspond to the (+) enantiomer and are illustrated in Fig.3.

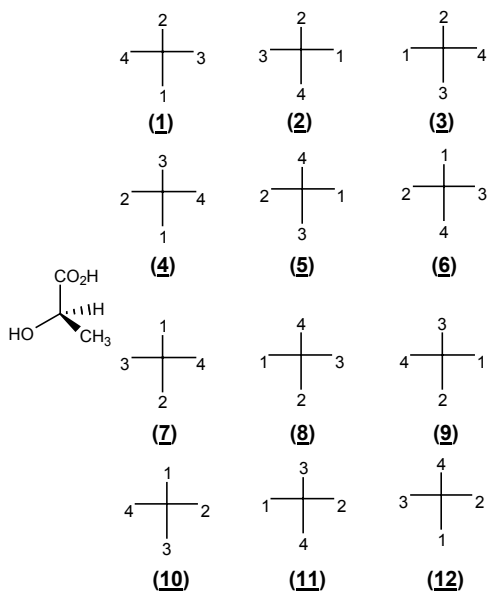

**Figure 4.** Twelve Fischer projections of (*R*)-(-)-lactic acid.

The other 12 graphs in Fig. 4 represent the (−) enantiomer.

The permutations shown in Fig.3 can be obtained, either by permuting groups of three bonds or by turning the projections by 180°. The permutations outlined in Fig.4 derive by those in Fig. 3 simply by interchanging two groups.

In order to reduce the Fischer rules to an algebraic structure, we define an operator $\chi_k$ acting on a tetrahedral molecule. Let us take into account a single tetrahedron. The generalization of the following results to simply connected chains of tetrahedrons is easily accomplished [4].

The Fischer projection corresponding to the column vector (27) is

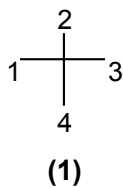

(1)

which is the first in Fig.3. The position of the bonds in the column vector (27) are assigned starting from the left and proceeding clockwise in the Fischer projection.

The matrix representation of the projection **(1)** is assumed as the starting one with respect to the operators $\chi_k$, i. e. $\chi_1$ is the 4 x 4 identity matrix

$$\chi_1 = \begin{pmatrix} 1 & 0 & 0 & 0 \\ 0 & 1 & 0 & 0 \\ 0 & 0 & 1 & 0 \\ 0 & 0 & 0 & 1 \end{pmatrix}, \tag{28}$$

so the action on the column vector $\Psi$ is

$$\chi_1 \begin{pmatrix} \Psi_1 \\ \Psi_2 \\ \Psi_3 \\ \Psi_4 \end{pmatrix} = \begin{pmatrix} \Psi_1 \\ \Psi_2 \\ \Psi_3 \\ \Psi_4 \end{pmatrix} \tag{29}$$

The configuration **(2)** of Fig.3 can be achieved as soon as we define an operator $\chi_2$ acting as

$$\chi_2 \begin{pmatrix} \Psi_1 \\ \Psi_2 \\ \Psi_3 \\ \Psi_4 \end{pmatrix} = \begin{pmatrix} \Psi_3 \\ \Psi_2 \\ \Psi_4 \\ \Psi_1 \end{pmatrix} \tag{30}$$

which corresponds to the matrix

$$\chi_2 = \begin{pmatrix} 0 & 0 & 1 & 0 \\ 0 & 1 & 0 & 0 \\ 0 & 0 & 0 & 1 \\ 1 & 0 & 0 & 0 \end{pmatrix} \tag{31}$$

It is clear that $\chi_2$ is a rotation. On the other hand, the configuration **(1)** of the (−) enantiomer can be obtained starting from the column vector (27), if we define an operator $\bar{\chi}_1$ which acts as

$$\bar{\chi}_1 \begin{pmatrix} \Psi_1 \\ \Psi_2 \\ \Psi_3 \\ \Psi_4 \end{pmatrix} = \begin{pmatrix} \Psi_4 \\ \Psi_2 \\ \Psi_3 \\ \Psi_1 \end{pmatrix} \tag{32}$$

Explicitly, we have

$$\bar{\chi}_1 = \begin{pmatrix} 0 & 0 & 0 & 1 \\ 0 & 1 & 0 & 0 \\ 0 & 0 & 1 & 0 \\ 1 & 0 & 0 & 0 \end{pmatrix} \tag{33}$$

It generates the inversion between the bonds $\Psi_1$ and $\Psi_4$.

By this approach, all the 24 projections can be obtained (12 for the (+) enantiomer and 12 for the (−) enantiomer represented in Figs.3 and 4) by matrix operators acting on the fundamental projection **(1)**. The tables in the Appendix summarize the situation. The matrices in Table I and II (see the Appendix) are the elements of a 4-parameter algebra. Those in Table I are a representation of rotations, while those in Table II are inversions. Both sets constitute the O(4) group of 4×4 orthogonal matrices. The matrices in Table I are a representation of the remarkable subgroup SO(4) of 4×4 matrices with determinant +1. The matrices in Table II have determinant −1, being inversions (or reflections). They do not constitute a group since the product of any two of them has determinant +1. This fact means that the product of two inversions generates a rotation (this is obvious by inverting both the couples of bonds in a tetrahedron). The total algebra is closed [9] and it represents all possible unitary transformations that a quaternion, in 4-vector form, can undergo.

The 24 matrices in Table I and II in the Appendix are not all independent. They can be grouped as different representations of the same operators. In fact all matrices, representing the same operator, have the same characteristic polynomial [9,11]. In other words, the characteristic equation of a matrix is invariant under vector base changes. In the (+) enantiomer case, the characteristic eigenvalue equation is

$$\det \| \chi_k - \lambda \mathbf{I} \| = 0 \tag{34}$$

where $\lambda$ is the eigenvalue and **I** is the identity matrix.

In the case of the (−) enantiomer, we have

$$\det \| \bar{\chi}_k - \lambda \mathbf{I} \| = 0 \tag{35}$$

From Eqs. (34) and (35), we recover 6 eigenvalues

$$\lambda_{1,2} = \pm 1 \quad \lambda_{3,4} = \pm i \quad \lambda_{5,6} = \frac{-1 \pm i\sqrt{3}}{2} \tag{36}$$

Inserting them into Eqs. (34) and (35), it is easy to determine the eigenvectors being

$$(\chi_k - \lambda \mathbf{I})\Psi = 0, \qquad (\bar{\chi}_k - \lambda \mathbf{I})\Psi = 0, \tag{37}$$

with obvious calculations depending on the choice of $\chi_k$ and $\bar{\chi}_k$. $\Psi$ is given by Eq. (27). It is worth noting that the number of independent eigenvalues (and then eigenvectors) is related to the number of independent elements of each member of a given group $O(N)$. If $N^2$ is the total number of elements and $\frac{1}{2}N(N+1)$ are the orthogonality conditions, we have

$$N^2 - \frac{1}{2}N(N+1) = \frac{1}{2}N(N-1) \tag{38}$$

For $O(4)$, it is 6, which is the number of independent generators of the group [9], giving the "dimension" of the group. With these considerations in mind, it can be stated that Fischer projections generates the algebraic structure of tetrahedral molecules and quaternions are a natural representation. In fact, considering the 4-vector (27) as a quaternion, it is clear that its components can undergo rotations and inversions satisfying the fundamental rules (2), (3), (4) (in particular any chiral transformation as (32)). Any single bond can be seen as a complex number [3] and then the considerations valid for the matrix (5) hold. In this sense, a couple of bonds satisfy the $SU(2)$ algebra, while the total tetrahedron satisfies the algebra (21), or $O(4)$. The matrices $\chi_k$ and $\bar{\chi}_k$ are unitary quaternionic operators accounting for all possible transformations preserving the tetrahedral structure. As a particular case, if we are considering an achiral tetrahedron, all the transformations preserving its structure can be accounted by the $SO(4)$ group, i. e. only rotations without inversions.

It is worthwhile noting that this approach, in particular Eq. (34), (35) and the eigenvalues (36), can be related to the "pseudoscalar measurements" used to get information about the structure (in particular the chirality) of molecules in the sense indicated by Ruch [12] and Barron [13]. In fact, a pseudoscalar, in particular its sign, can always be related to optical rotation or circular dichroism allowing the determination of the molecular chirality. The double sign of eigenvalues (36) is an indication in this sense, because it clearly points out the fact that a molecule is passing from a state with a given chirality to the opposite one.

However, our $\lambda_s$ are not the Ruch's $\lambda_s$. In our case, $\lambda$ indicates the global chirality state of the molecule, in Ruch's case, it is a parameter describing the ligand. The stereochemical information, in our case, is relative to both enantiomers and it is not absolute. The transitions between two chiral quantum states of the molecules are discuss in the next section.

## 4. Quaternionic algebra and quantum states of molecules

In order to realize the Heisenberg program, our quaternionic algebra has to account for the transitions of quantum states of molecules. In fact, always along the line suggested by Heisenberg, a crucial point is that elementary particles are not in stationary states of the hamiltonian and so their quantum states do not belong to definite symmetries. This is also the case for chiral molecules, which interconvert (via tunnelling) between the left- and right-handed states, which therefore do not have definite parity (since parity interconverts the two states).

In order to study the dynamics of these transitions, we can take into account the set of operators $\chi_k$, $\bar{\chi}_k$ and $\hat{\mathcal{H}}^{tot}$, where $\chi_k$, $\bar{\chi}_k$ are the above unitary quaternionic operators, which give all the possible transformations preserving the tetrahedral structure, and $\hat{\mathcal{H}}^{tot}$ is the total Hamiltonian operator for the degenerate isomers of an optically active molecule which has to consists of an even and an odd part [14], i.e.

$$\hat{\mathcal{H}}^{tot} = \hat{\mathcal{H}}^{even} + \hat{\mathcal{H}}^{odd} \qquad (39)$$

If this is the case, ignoring eventual parity violation effects [15], energy eigenstates can be obtained as superpositions of handed states, that is, by using the Dirac ket notation,

$$|\Psi_\pm\rangle = \frac{1}{\sqrt{2}}(|\Psi_L\rangle \pm |\Psi_R\rangle). \qquad (40)$$

These assumptions guarantee that the Hund result

$$\left[\hat{P}, \hat{\mathcal{H}}^{tot}\right] = 0 \qquad (41)$$

holds. $\hat{P}$ is the parity operator whose eigenvalues are p= ±1. The unitary quaternionic operators $\chi_k$ and $\bar{\chi}_k$ act on the handed states in the superposition (40). In fact, starting from the initial state (27), which we will define as the initial configuration, we have

$$|\Psi_R\rangle_k = \chi_k |\Psi\rangle_0 \tag{42}$$

and

$$|\Psi_L\rangle_k = \bar{\chi}_k |\Psi\rangle_0 \tag{43}$$

which means that, considering the quaternionic operators in Tables I and II, the operators $\chi_k$ "rotate" the molecule while the operators $\bar{\chi}_k$ " invert" a couple of bonds. Dropping the indexes and considering any state of the molecule (not necessarily the initial configuration (27)), we have four possibilities

$$\chi|\Psi_R\rangle = |\Psi_R\rangle; \quad \chi|\Psi_L\rangle = |\Psi_L\rangle \tag{44}$$

$$\bar{\chi}|\Psi_R\rangle = |\Psi_L\rangle; \quad \bar{\chi}|\Psi_L\rangle = |\Psi_R\rangle \tag{45}$$

where the $\bar{\chi}_k$ operators interconvert the two handed nonstationary states acting as an algebraic counterpart of quantum tunnelling. Considering the energy eigenstates (40), which, thanks to (41) are also parity eigenstates, we have

$$\chi|\Psi_\pm\rangle = \frac{1}{\sqrt{2}}\left(\chi|\Psi_L\rangle \pm \chi|\Psi_R\rangle\right) \tag{46}$$

$$= \frac{1}{\sqrt{2}}\left(|\Psi_L\rangle \pm |\Psi_R\rangle\right)$$

and

$$\bar{\chi}|\Psi_\pm\rangle = \frac{1}{\sqrt{2}}\left(\bar{\chi}|\Psi_L\rangle \pm \bar{\chi}|\Psi_R\rangle\right) \tag{47}$$

$$= \frac{1}{\sqrt{2}}\left(|\Psi_R\rangle \pm |\Psi_L\rangle\right).$$

This fact means that the $\bar{\chi}_k$ operators allow the transition between the $|\Psi_R\rangle$-states and the $|\Psi_L\rangle$-states and viceversa (as a quantum tunnelling process), but the overall parity of the energy state is preserved. In conclusion, the unitary quaternionic operators $\chi_k$ and $\bar{\chi}_k$, accounting for the transformations preserving the tetrahedral structure, allow the transitions between the non-stationary chiral states of the molecule: specifically the $\chi$ operators preserve the given handed state only rotating it, the $\bar{\chi}$ operators

interconvert the handed states inverting a couple of bonds. Both of them are quaternionic transformations.

## 5. Discussion and conclusions

In our opinion, the above considerations could have a deep meaning in the footpath indicated by Heisenberg [1]. In fact, particles as Dirac spinors can be described exactly by the same approach by using unitary quaternion operators which act on quantum states.

As an example, let us take into account a free particle $\Psi$ whose dynamics is described by the Klein-Gordon equation

$$\frac{1}{c^2}\frac{\partial^2\Psi}{\partial t^2} = \frac{\partial^2\Psi}{\partial x^2} + \frac{\partial^2\Psi}{\partial y^2} + \frac{\partial^2\Psi}{\partial z^2} - \frac{m^2 c^2}{\hbar^2}\Psi, \qquad (48)$$

We use the approach and the conventions formerly used by Lanczos [6] and Conway [7].

Thanks to the property (3) of quaternions, we can define a set of quaternionic coordinates so that

$$e_1^2 = e_2^2 = e_3^2 = i^2 = -1, \qquad (49)$$

by which the Klein-Gordon operator can be recast as

$$\frac{1}{c^2}\frac{\partial^2\Psi}{\partial t^2} = e_3^2\frac{\partial^2\Psi}{\partial x^2}e_2^2 + ie_1^2\frac{\partial^2\Psi}{\partial y^2}i + e_3^2\frac{\partial^2\Psi}{\partial z^2}e_1^2 + i^2\frac{m^2c^2}{\hbar^2}e_3^2\Psi e_3^2. \qquad (50)$$

Clearly m is the mass of the particle, c is the speed of light and $\hbar$ is the Planck constant. Eq. (50) can be obtained if $\Psi$ is a quaternionic 4-vector particle on which the operator

$$\frac{1}{c}\frac{\partial}{\partial t} = e_3\frac{\partial}{\partial x}e_2 + e_1\frac{\partial}{\partial y}i + e_3\frac{\partial}{\partial z}e_1 + i\frac{mc}{\hbar}e_3(\ )e_3 \qquad (51)$$

is applied twice. The operators in the r.h.s. of Eq. (51)

$$e_3(\ )e_2, \quad e_1(\ )i, \quad e_3(\ )e_1, \qquad (52)$$

are the Conway matrices which can be recast in the compact form $-\alpha_k$, that is

$$\alpha_k = \begin{pmatrix} 0 & \sigma_k \\ \sigma_k & 0 \end{pmatrix} \qquad (53)$$

where $\sigma_k$ are the three Pauli matrices (9), while the Conway matrix

$$e_3(\ )e_3 = \begin{pmatrix} 1 & 0 & 0 & 0 \\ 0 & 1 & 0 & 0 \\ 0 & 0 & -1 & 0 \\ 0 & 0 & 0 & -1 \end{pmatrix} \quad (54)$$

is the well-known Dirac $\beta$ matrix. Through these notations, the operator (51) gives the Dirac equation for a free fermion

$$\frac{1}{c}\frac{\partial \Psi}{\partial t} = -\alpha_k \frac{\partial \Psi}{\partial x^k} - i\frac{mc}{\hbar}\beta\Psi \quad (55)$$

in a non-covariant form. The covariant form of Dirac equation [8] is immediately recovered applying the operator $ie_3(\ )e_3$ to Eq. (55). The isomorphism between quaternionic and spinorial algebrae, through the multiplication rules of Pauli matrices, gives the identification of the particle wavefunction $\Psi$ with a Dirac spinor. In other words, spinors and quaternions can be represented by the same groups [11] (e.g. SL(2,C) or O(4)) as tetrahedral molecules.

An immediate objection to the approach which considers particles and molecules under the same standard could be that particles are relativistic objects, while molecules are not, so the analogy could be only formal. This is not true since in the limit $v \ll c$, the Dirac equation becomes the Pauli equation [16] which is nothing else but a Schrödinger equation for spinors where an additional term of magnetic dipole has to be consider. In this perspective, dynamics and algebra of tetrahedrons and particles are the same and the differences lie only in energy, length and time scales. In other words, quantum mechanics of spinor particles and chiral molecules is fundamentally scale invariant.

As a remark, it is interesting to stress the fact that the chirality of tetrahedrons and spinor particles is recovered by quaternionic approach in an abstract 4-space thanks to the intrinsic anticommutativity of the algebraic representation as $O(4)$ or $SU(2)\times SU(2)$. It is worth noting that the dimensionality of the space plays a fundamental role in defining the chirality. As a final remark, it is worth pointing out that Fischer projections, used by organic chemists to determine absolute configurations of optically pure molecules or simply to represent them bidimensionally, seem not only an empirical set of rules, but can constitute a fundamental structure to achieve a quantum chiral algebra, which relates properties of tetrahedral molecules to those of spinor particles.

## Acknoledgements

The authors thank the referee for the useful suggestions and comments which allowed to improve the paper.

**Appendix**

The operators $\chi_k$ give rise to the representations of the (+) enantiomer, while the operators $\bar{\chi}_k$ give rise to those of the (−) enantiomer. Obviously $k =1,..,12$.

Table I, (+)-enantiomer:

$$\chi_1 = \begin{pmatrix} 1 & 0 & 0 & 0 \\ 0 & 1 & 0 & 0 \\ 0 & 0 & 1 & 0 \\ 0 & 0 & 0 & 1 \end{pmatrix} \quad \chi_2 = \begin{pmatrix} 0 & 0 & 1 & 0 \\ 0 & 1 & 0 & 0 \\ 0 & 0 & 0 & 1 \\ 1 & 0 & 0 & 0 \end{pmatrix} \quad \chi_3 = \begin{pmatrix} 0 & 0 & 0 & 1 \\ 0 & 1 & 0 & 0 \\ 1 & 0 & 0 & 0 \\ 0 & 0 & 1 & 0 \end{pmatrix}$$

$$\chi_4 = \begin{pmatrix} 0 & 1 & 0 & 0 \\ 0 & 0 & 1 & 0 \\ 1 & 0 & 0 & 0 \\ 0 & 0 & 0 & 1 \end{pmatrix} \quad \chi_5 = \begin{pmatrix} 0 & 1 & 0 & 0 \\ 1 & 0 & 0 & 0 \\ 0 & 0 & 0 & 1 \\ 0 & 0 & 1 & 0 \end{pmatrix} \quad \chi_6 = \begin{pmatrix} 0 & 1 & 0 & 0 \\ 0 & 0 & 0 & 1 \\ 0 & 0 & 1 & 0 \\ 1 & 0 & 0 & 0 \end{pmatrix}$$

$$\chi_7 = \begin{pmatrix} 0 & 0 & 1 & 0 \\ 0 & 0 & 0 & 1 \\ 1 & 0 & 0 & 0 \\ 0 & 1 & 0 & 0 \end{pmatrix} \quad \chi_8 = \begin{pmatrix} 0 & 0 & 0 & 1 \\ 1 & 0 & 0 & 0 \\ 0 & 0 & 1 & 0 \\ 0 & 1 & 0 & 0 \end{pmatrix} \quad \chi_9 = \begin{pmatrix} 1 & 0 & 0 & 0 \\ 0 & 0 & 1 & 0 \\ 0 & 0 & 0 & 1 \\ 0 & 1 & 0 & 0 \end{pmatrix}$$

$$\chi_{10} = \begin{pmatrix} 1 & 0 & 0 & 0 \\ 0 & 0 & 0 & 1 \\ 0 & 1 & 0 & 0 \\ 0 & 0 & 1 & 0 \end{pmatrix} \quad \chi_{11} = \begin{pmatrix} 0 & 0 & 0 & 1 \\ 0 & 0 & 1 & 0 \\ 0 & 1 & 0 & 0 \\ 1 & 0 & 0 & 0 \end{pmatrix} \quad \chi_{12} = \begin{pmatrix} 0 & 0 & 1 & 0 \\ 1 & 0 & 0 & 0 \\ 0 & 1 & 0 & 0 \\ 0 & 0 & 0 & 1 \end{pmatrix}$$

Table II, (−) enantiomer:

$$\bar{\chi}_1 = \begin{pmatrix} 0 & 0 & 0 & 1 \\ 0 & 1 & 0 & 0 \\ 0 & 0 & 1 & 0 \\ 1 & 0 & 0 & 0 \end{pmatrix} \quad \bar{\chi}_2 = \begin{pmatrix} 0 & 0 & 1 & 0 \\ 0 & 1 & 0 & 0 \\ 1 & 0 & 0 & 0 \\ 0 & 0 & 0 & 1 \end{pmatrix} \quad \bar{\chi}_3 = \begin{pmatrix} 1 & 0 & 0 & 0 \\ 0 & 1 & 0 & 0 \\ 0 & 0 & 0 & 1 \\ 0 & 0 & 1 & 0 \end{pmatrix}$$

$$\bar{\chi}_4 = \begin{pmatrix} 0 & 1 & 0 & 0 \\ 0 & 0 & 1 & 0 \\ 0 & 0 & 0 & 1 \\ 1 & 0 & 0 & 0 \end{pmatrix} \quad \bar{\chi}_5 = \begin{pmatrix} 0 & 1 & 0 & 0 \\ 0 & 0 & 0 & 1 \\ 1 & 0 & 0 & 0 \\ 0 & 0 & 1 & 0 \end{pmatrix} \quad \bar{\chi}_6 = \begin{pmatrix} 0 & 1 & 0 & 0 \\ 1 & 0 & 0 & 0 \\ 0 & 0 & 1 & 0 \\ 0 & 0 & 0 & 1 \end{pmatrix}$$

$$\bar{\chi}_7 = \begin{pmatrix} 0 & 0 & 1 & 0 \\ 1 & 0 & 0 & 0 \\ 0 & 0 & 0 & 1 \\ 0 & 1 & 0 & 0 \end{pmatrix} \quad \bar{\chi}_8 = \begin{pmatrix} 1 & 0 & 0 & 0 \\ 0 & 0 & 0 & 1 \\ 0 & 0 & 1 & 0 \\ 0 & 1 & 0 & 0 \end{pmatrix} \quad \bar{\chi}_9 = \begin{pmatrix} 0 & 0 & 0 & 1 \\ 0 & 0 & 1 & 0 \\ 1 & 0 & 0 & 0 \\ 0 & 1 & 0 & 0 \end{pmatrix}$$

$$\bar{\chi}_{10} = \begin{pmatrix} 0 & 0 & 0 & 1 \\ 1 & 0 & 0 & 0 \\ 0 & 1 & 0 & 0 \\ 0 & 0 & 1 & 0 \end{pmatrix} \quad \bar{\chi}_{11} = \begin{pmatrix} 1 & 0 & 0 & 0 \\ 0 & 0 & 1 & 0 \\ 0 & 1 & 0 & 0 \\ 0 & 0 & 0 & 1 \end{pmatrix} \quad \bar{\chi}_{12} = \begin{pmatrix} 0 & 0 & 1 & 0 \\ 0 & 0 & 0 & 1 \\ 0 & 1 & 0 & 0 \\ 1 & 0 & 0 & 0 \end{pmatrix}$$